# PRK AND ROCK ARE REQUIRED FOR THROMBIN-INDUCED ENDOTHELIAL CELL PERMEABILITY DOWNSTREAM FROM $G\alpha_{12/13}$ and $G\alpha_{11/q}$

Julie Gavard[1,*] and J Silvio Gutkind[1]
From [1] Oral and Pharyngeal Cancer Branch, National Institute of Dental and Craniofacial Research, National Institutes of Health, Bethesda, Maryland 20892
Running head: RhoA bifurcation on ROCK and PRK
*Address correspondance to: Julie Gavard, Ph.D., CNRS, UMR8104, INSERM, U567, Université Paris Descartes, Institut Cochin, 22 rue Méchain, 75014 Paris, France. Fax: +33140516430. Email: julie.gavard@inserm.fr

**Increase in vascular permeability occurs under many physiological conditions such as wound repair, inflammation, and thrombotic reactions, and is central in diverse human pathologies, including tumor-induced angiogenesis, ocular diseases and septic shock. Thrombin is a pro-coagulant serine protease, which causes the local loss of endothelial barrier integrity thereby enabling the rapid extravasation of plasma proteins and the local formation of fibrin-containing clots. Available information suggests that thrombin induces endothelial permeability by promoting acto-myosin contractility through the Rho/ROCK signaling pathway. Here, we took advantage of pharmacological inhibitors, knock-down approaches, and the emerging knowledge on how permeability factors affect endothelial junctions to investigate in detail the mechanism underlying thrombin-induced endothelial permeability. We show that thrombin signals through PAR-1 and its coupled G proteins $G\alpha_{12/13}$ and $G\alpha_{11/q}$ to induce RhoA activation and intracellular calcium elevation, and that these events are interrelated. In turn, this leads to the stimulation of ROCK, which causes actin stress-fiber formation. However, this alone is not sufficient to account for thrombin-induced permeability. Instead, we found that PRK, a Rho-dependent serine/threonine kinase, is activated in endothelial cells upon thrombin stimulation, and that its expression is required for endothelial permeability and the remodeling of cell-extracellular matrix and cell-cell adhesions. Our results demonstrate that the signal initiated by thrombin bifurcates at the level of RhoA, to promote changes in the cytoskeletal architecture through ROCK, and the remodeling of focal adhesion components through PRK. Ultimately, both pathways converge to cause cell-cell junction disruption and provoke vascular leakage.**

Endothelial homeostasis and vascular integrity are tightly regulated during normal angiogenesis, wound repair, thrombotic and inflammatory reactions (1). The vascular wall controls the exchange of macromolecules and fluid between blood compartment and interstitial tissue (2). Whereas pro-angiogenic pathways had been extensively studied, as part of efforts to understand normal and aberrant angiogenesis, the molecular mechanisms involved in the vascular barrier permeability, and their implications in aberrant angiogenesis are still much less understood. For example, VEGF (Vascular endothelial growth factor), first described as Vascular Permeability Factor (VPF) acts by a biochemical route that involve the sequential activation of VEGF-receptor 2 (VEGFR2), the kinase Src, the guanine exchange factor Vav2, and the GTPase Rac and p21 activated-kinase PAK (3-5). This pathway converges on the regulation of endothelial cell-cell junctions thereby causing their disruption by promoting the internalization of the endothelial adherens junction protein, VE-cadherin (3,6,7).

The coagulation protease, thrombin, which activates the PAR (protease-activated receptor) family of G-protein coupled receptor (8) by proteolytic cleavage, represents another key regulator of the endothelial barrier function. It is well known that endothelial exposure to thrombin stimulation induces rapid morphological and cytoskeletal changes, characterized by formation of actin stress fibers and endothelial gaps that could both be involved in the loss of endothelial barrier integrity (9). Several studies support that RhoA activation downstream $G\alpha_{12/13}$ coupled to PAR-1 is required for these cellular events (10-12). In addition, ROCK, myosin light chain (MLC) and actin-regulating proteins participate in thrombin-triggered cytoskeletal reorganization and endothelial barrier disruption, most likely through the acto-myosin contractility pathway (13-15).

However, the intervening molecular mechanisms are probably more complex than this linear biochemical route, as $G\alpha_{11/q}$ coupling and calcium signaling, as well as the calcium-activated kinases PKC, have also been shown to play a role in thrombin-induced actin stress fiber formation (16-22). In addition, microtubule stability may also participate in cell contractility (23,24). Finally, RhoA activation may utilize downstream targets in addition to ROCK as, for example, some formin family proteins have been shown to contribute to thrombin-based endothelial cytoskeleton rearrangement (25,26). We therefore decided to investigate the thrombin-initiated molecular cascade leading to increased endothelial permeability using a siRNA-based knockdown approach for key signaling candidates. Here, we show that thrombin uses a bipartite coupling from PAR-1 involving both $G\alpha_{11/q}$ and $G\alpha_{12/13}$, which causes RhoA activation. The signal downstream from RhoA in turn bifurcates to stimulate two serine/threonine kinases, ROCK and PRK (PKC-related kinase). These RhoA downstream effectors then contribute to acto-myosin cell contractility, by regulating actin stress fiber formation and focal adhesion organization, respectively. Finally, these pathways converge to promote the redistribution of endothelial cell-cell junctions, and the disruption of VE-cadherin adhesion by a mechanism distinct from that triggered by VEGF stimulation. These findings may help to dissect the molecular mechanisms deployed by thrombin causing the loss of endothelial barrier integrity, which may facilitate the future development of anti-permeability agents in thrombotic reactions.

## EXPERIMENTAL PROCEDURES

*Cell culture and transfections*—Immortalized human vascular endothelial cells were obtained from (27), and cultured in Dulbecco's modified Eagle's medium supplemented with 10% fetal bovine serum (Sigma, Saint-Louis, MO). DNA transfections of pCEFL-mock, pCEFL-$G\alpha_{11}$, -$G\alpha_{12}$, -$G\alpha_{13}$, -$G\alpha_q$ wild-type or activated QL mutants (21), pCEFL-ROCK-KD (kinase dead) mutant (28) were performed using the Amaxa's electroporation system (Amaxa Biosystems, Gaithersburg, MD). Non-silencing control RNA sequences (nsi) and $G\alpha_{11}$, $G\alpha_{12}$, $G\alpha_{13}$, $G\alpha_q$ and PRK1 silencing RNA duplexes (siRNA, Qiagen, Valencia, CA) were transfected twice, on day 1 and day 3, using the Hiperfect reagent (Qiagen) and harvested after 5 days.

*Reagents and antibodies*—Recombinant human VEGF and IL-8 were purchased from PeproTech (Rocky Hill, NJ), thrombin was purchased from Sigma, as well as ionomycin, BAPTA-AM (1,2-Bis 2-aminophenoxy ethane-N,N,N′,N′-tetraacetic acid tetrakis acetoxymethyl ester), 2-ABP (2-amino-1-hydroxybutylidene-1,1-bisphosphonic acid), TMB-8 (3,4,5-trimethoxybenzoic acid 8-diethylamino-octyl ester), lanthanade, verapamil (5-N-3,4-dimethoxyphenylethylmethylamino-2-3,4-dimethoxyphenyl-2-isopropylvaleronitrile hydrochloride), and ML-7 (1-5-iodonaphthalene 1-sulfonyl 1H-hexahydro 1,4-diazepine hydrochloride). Pertussis toxin, botulinum coenzyme C3 toxin and tetanolysin were from List Biologicals Labs (Campbell, CA), blebbistatin (1-Phenyl 1,2,3,4-tetrahydro 4-hydroxypyrrolo 2.3-b-7-methylquinolin 4-one) and Y27632 from Calbiochem (San Diego, CA). SCH79797 Par-1 and TCY-NH2 Par-3/4 inhibitors were from Tocris (Ellisville, MO), FSLLRY-NH2 Par-2 from Peptides International (Louisville, KY). The following antibodies were used: anti-paxillin (BD Biosciences, San Jose, CA), anti-phospho-extracellular regulated kinases ERK1/2, phospho-Y118 paxillin, phospho-S19 Myosin Light Chain (MLC), phospho-T774/T816 PRK1/2, and phospho-S3 cofilin (Cell Signaling, Boston, MA), anti-myc (Covance, Princeton, NJ), anti-phospho-tyrosine (4G10, Millipore, Waltham, MA), anti-phospho-Y397 Focal Adhesion Kinase FAK (BioSource, Invitrogen, Camarillo, CA), anti-ZO-1 (Zymed, Invitrogen), anti-tubulin, extracellular related kinase ERK1, $G\alpha_{11}$, $G\alpha_{12}$, $G\alpha_{13}$, $G\alpha_q$, RhoA, PRK1 and VE-cadherin (Santa Cruz Biotech, Santa Cruz, CA), anti-β-catenin (Sigma), anti-phospho-S665 VE-cadherin (3), and anti-β-arrestin2 (29).

*Cytosolic calcium measurements*—Changes in intracellular $Ca^{2+}$ levels were assessed using Fluo-4 NW calcium indicators from Invitrogen, following the manufacturer's instructions. Fluorescence was read in 96-well plate format at room temperature, using appropriate settings (Ex/Em 494/516 nm, Optima BMG Labtech, Durham, NC).

*Permeability assays*—*In vitro* permeability assays were conducted as described in (30), using 3 days-old endothelial monolayers on collagen-coated 3 μm porous membranes (Transwell, Corning Costar, Acton, MA) and 60 kDa fluorescein-conjugated dextran (Invitrogen). Measurements were analyzed on a multi-plate fluorescent reader (Victor 3V1420, Perkin-Elmer, Wellesley, PA).

*GST-pull downs, immunoprecipitations and western-blot*—Methods and buffers were all described in (30). Membranes were scanned using the Odyssey Infra-Red Imaging System from Li-Cor BioSciences (Lincoln, NB). Alexa fluorescent dye-conjugated secondary antibodies were purchased from Invitrogen.

*VE-cadherin internalization assay and immunofluorescence*—VE-cadherin antibody uptake and immunofluorescence staining protocols were described in (3,31). Secondary antibodies were from Jackson ImmunoResearch (West Grove, PA) and Alexa546-conjugated phalloidin from Invitrogen. All confocal acquisitions were performed on a TCS/SP2 Leica microscope (NIDCR Confocal Facility, Bethesda, MD).

*Statistical analysis*—Graphs are shown as a mean value ± SEM from at least 3 independent experiments, confocal pictures and western-blot scans are representative of at least 3 independent experiments. Statistical analysis was performed with the Prism software (ANOVA test, GraphPad), *** $p<0.001$, ** $p<0.01$, * $p<0.05$.

## RESULTS

*Thrombin induces endothelial permeability mainly through PAR-1 and $G\alpha_{12/13/11/q}$ coupling*—Since multiple receptors for thrombin (PAR) might co-exist at the surface of endothelial cells, we first assessed which PAR was involved during the endothelial permeability response using competitive antagonists. Whereas blocking PAR-2, PAR-3 and PAR-4 only mildly affected thrombin-induced permeability, interfering with PAR-1 signaling dramatically reduced it to its basal level (Fig. 1A). Interestingly, inhibition of ERK activation was achieved only when multiple PAR antagonists were used, suggesting that whereas several PARs can independently controlled ERK activation in endothelial cells, PAR-1 may play a more prominent role in initiating signaling events culminating in endothelial barrier disruption (Fig. 1B). We then aimed at characterizing the Gα subunits responsible for the PAR-1-controlled permeability. Initially, by the use of pertussis toxin which impedes $G\alpha_i$ signaling (32), we confirmed that thrombin-induced permeability was independent of $G\alpha_i$, in contrast to IL-8, which acts through its cognate $G\alpha_i$-coupled receptor, CXCR2 (Fig. 1C). To determine whether $G\alpha_{12/13}$ or $G\alpha_{11/q}$ subunits are instead involved, we first utilized a molecular mutant approach, where dominant active forms were transfected in endothelial cells (Fig. 1D). Interestingly, the expression of activated forms of $G\alpha_q$ and $G\alpha_{11}$ was sufficient to increase endothelial permeability in non-stimulated cells (Fig. 1E). However, all $G\alpha_{12}$, $G\alpha_{13}$, $G\alpha_{11}$ and $G\alpha_q$ dominant active forms were able to potentiate thrombin-induced junctional permeability (Fig. 1F), suggesting a complex coupling of PAR receptors in the endothelial barrier function. Importantly, they did not alter the response to VEGF, a pro-permeability factor that acts through a tyrosine-kinase receptor (Fig. 1F). To understand better the specificity of PAR receptors signaling in endothelial permeability regulation, we next investigate the contribution of each Gα subunits by siRNA-based knockdown approaches (Fig. 1G). This revealed that $G\alpha_{11}$ and $G\alpha_q$ were both required for thrombin to increase endothelial permeability, while $G\alpha_{12}$ and $G\alpha_{13}$ only partially affected the amplitude of the permeability response. Thus, our data suggest that thrombin can increase endothelial permeability mainly through PAR-1 coupled to $G\alpha_{11/q}$, and to a lesser extent through $G\alpha_{12/13}$.

*$G\alpha_{11/q}$-induced calcium entry is required for thrombin signaling in permeability*—$G\alpha_{11/q}$-coupled receptors can generally increase intracellular calcium concentrations, thus triggering multiple signaling pathways. Indeed, as previously reported, thrombin provoked a rapid increase in intracellular $Ca^{2+}$ concentration (Fig. 2A)(16). This calcium wave was strictly dependent on the presence of $G\alpha_{11/q}$, but not of $G\alpha_{12/13}$, as judged by the calcium flux in cells in which specific Gα subunits was knocked down, alone or in combination (Fig. 2B). We thus explored the role of calcium entry in endothelial permeability. We first used different chemical inhibitors with distinct mode of action. BAPTA-AM, a general divalent ion chelator, reduced the increase of endothelial permeability upon thrombin stimulation but also in response to VEGF, suggesting a broad-spectrum inhibition of essential calcium-dependent events (Fig. 1C). Interestingly 2-ABP and TMB-8, two different compounds blocking calcium release from intracellular compartment storage (33), did not modify endothelial permeability, whereas lanthanade and verapamil, which both affect extracellular $Ca^{2+}$ influx through plasma membrane receptor-operated calcium channels (16), dramatically prevented thrombin-induced permeability (Fig. 1C). Thus, these data suggest that thrombin can trigger an increase in intracellular calcium concentration by the activation of receptor-operated calcium channels on endothelial cells, which is in turn necessary for the endothelial barrier control. As it is well described that vascular permeability increase by thrombin is mediated by RhoA activation (9,16), we asked whether calcium increase and RhoA activity were linked. Interestingly, we first observed that in endothelial cells both G$\alpha_{12/13}$ and G$\alpha_{11/q}$ protein subunit family contribute to RhoA activation by thrombin (Fig. 2D). Moreover, when the increase in intracellular calcium concentration was halted by the use of lanthanade or chelated by BAPTA-AM, thrombin-induced RhoA activation was barely detectable (Fig. 2E). Instead, 2-ABP, which affects storage-operated calcium, did not alter RhoA activation, in agreement with its lack of effect on endothelial permeability. In line with these results, myosin light chain (MLC) phosphorylation, a downstream target of RhoA, was similarly prevented by BAPTA-AM and lanthanade (Fig. 2E). Together, these findings suggest that thrombin stimulates G$\alpha_{11/q}$ to promote calcium entry, which in turn cooperates with G$\alpha_{12/13}$ to stimulate RhoA activation and endothelial permeability.

*ROCK and PRK are both required for thrombin-induced permeability*—Since RhoA is an essential mediator for thrombin-triggered permeability through both G$\alpha_{12/13}$ and G$\alpha_{11/q}$, we examined in more detail the mechanisms involved in endothelial permeability downstream from RhoA. First, we observed that the inhibition of RhoA activation by C3 toxin blocked thrombin-induced permeability, while VEGF function was not affected (Fig. 3A). In addition, the use of Y27632, a well characterized inhibitor of ROCK and PRK (34), two serine/threonine kinases acting downstream of RhoA, hampered the thrombin-induced permeability response (Fig. 3A). The chemical inhibition of RhoA-dependent cytoskeletal contractility by ML7 and blebbistatin, which inhibit MLC kinase and myosin II, respectively, also prevented the increase in endothelial permeability caused by thrombin, while mildly affecting the effect of VEGF (Fig. 3B). These observations are aligned with the role of cell contractility in endothelial permeability upon thrombin activity (9). To assess the specific contribution of ROCK and PRK in this RhoA-initiated contractility-dependent pathway, we decided to knockdown PRK expression by the use of siRNA and to abolish ROCK activity by the expression of its dominant-negative mutant (ROCK-KD). The latter was found to be a more effective approach to prevent ROCK function than knocking down ROCK, which was only limited using multiple siRNAs (data not shown). As reported earlier, ROCK is required for thrombin-dependent permeability increase (Fig. 3C)(14). Surprisingly, we found that PRK is also necessary for this response, as siRNA extinction of PRK expression prevented from thrombin-induced permeability (Fig. 3C). Thus, both RhoA downstream targets, ROCK and PRK are required for the endothelial barrier destabilization caused by thrombin exposure.

*ROCK and PRK control separately thrombin-induced contractility pathways*—We then aimed at understand better the specific contribution of each RhoA target in the regulation of the endothelial barrier by thrombin. Among the many mediators of cell contractility, actin polymerization, stress fiber formation and focal adhesion attachment play essential roles in endothelial cell function (35). Thus, we explored the impact of PRK and ROCK on these multiple cellular events. First, in the absence of PRK expression, we observed that thrombin-induced phosphorylation of cofilin and MLC were not altered, while FAK and paxillin phosphorylations were severely reduced (Fig. 4A-C). These data indicated that PRK is essential for thrombin-regulated phosphorylation of focal adhesion components. On the contrary, blockade of ROCK activity impeded cofilin and MLC phosphorylation, which mediate actin polymerization, but leaving intact the levels of activation of paxillin and FAK (Fig. 4B-C). To further estimate the implication of ROCK and PRK on the contractility pathways at the level of actin stress fiber and focal adhesion formation, endothelial cells were prepared for immunostaining as described in (36). First, actin staining in endothelial monolayers showed a regular pattern of actin aligned at cell-cell contact area. Thrombin induced a massive formation of actin stress fibers, and opening of the cell-cell junctions in agreement with its effect on endothelial permeability (Fig. 4D). However, PRK siRNA and ROCK-KD expression altered these morphological changes, as shown by actin labeling with phalloidin and the quantification of the number of actin stress fibers (Fig. 4E). Similarly, we analyzed the status of focal adhesion by paxillin staining. Interestingly, the number of paxillin clusters was reduced in PRK siRNA-treated cells (Fig. 4F). While found aligned to sparse actin fibers perpendicular to the cell-cell contact borders, the morphology of paxillin staining was dramatically altered, including their size, suggesting an abortive attachment of endothelial cells to the extracellular matrix in the absence of PRK (Fig. 4G). Our observations further support the emerging role of PRK in the assembly and function of focal adhesions in endothelial cells exposed to thrombin. Conversely, ROCK-KD expression severely blocked the formation of actin cables crossing the cell (Fig. 4D-E), while the morphology and the number of the paxillin clusters were barely modified (Fig. 4F-G). Altogether, our data support the notion of a two-pronged regulation of acto-myosin contractility in response to thrombin exposure. In this model, RhoA stimulation triggers, PRK activation, which regulates focal adhesion components, and ROCK, which acts though MLC and cofilin to induce actin stress fiber formation. Both events ultimately cooperate to promote endothelial cell contractility and junctional permeability.

*ROCK and PRK-dependent contractility is involved in cell-cell contact remodeling*—VEGF-induced permeability requires cell-cell junction remodeling (3,6,7). Our results prompted us to characterize how cell contractility could control junctional permeability. First, we observed that thrombin stimulation was associated with the reorganization of VE-cadherin and ZO-1-containing contacts (Fig. 5A). When analyzed in more detail, we noticed that the VE-cadherin staining-containing areas appeared larger upon VEGF stimulation, which can reflect overlapping of membranes, ruffling or intracellular accumulation at the vicinity of cell-cell contacts. Thrombin exposure resulted in a slightly different morphology, as VE-cadherin staining looked more stretched as the cell-cell junctions were disrupted (Fig. 5A). We can speculate that thrombin may use a contractility-based mechanism that tears apart cell-cell junctions, whereas VEGF may deploy signaling pathways resulting in a more subtle effect, involving VE-cadherin redistribution and membrane remodeling (3,5,37,38). In this context, VE-cadherin endocytosis was markedly increased upon VEGF exposure while enhanced by a fold above the basal level by thrombin (Fig. 5B). In addition, while both permeability factors led to tyrosine phosphorylation of VE-cadherin and reduction of VE-cadherin/β-catenin interaction, only VEGF promoted S665 phosphorylation of VE-cadherin and its further interaction with β-arrestin (Fig. 5C), as a hallmark for VE-cadherin adhesion internalization (30,39,40). Finally, both ROCK and PRK blockade reduced the effects of thrombin on VE-cadherin junctions, as the overall organization of the endothelial monolayer appeared less disrupted (Fig. 5D). These observations suggest that VEGF and thrombin trigger divergent molecular mechanisms that may converge on the destabilization of endothelial cell-cell junctions and loss of endothelial monolayer integrity.

## DISCUSSION

Thrombin, a key pro-coagulant serine-protease, also regulates the adhesion and rolling of immune cells on the luminal face of blood vessel walls, and exerts a potent pro-angiogenic and pro-permeability action through endothelial expressed PARs. These G-protein coupled receptors involve the activation of the small GTPase RhoA in endothelial cells in response to thrombin (9,41). However, the intervening G protein subunits and the precise RhoA downstream targets by which thrombin acts are much less understood. Here, we show that thrombin-induced RhoA activation is coupled to $G\alpha_{12/13}$ and to $G\alpha_{11/q}$, and that the latter involves intracellular calcium concentration elevation. Furthermore, we show that the signaling pathway initiated by thrombin bifurcates downstream from RhoA, as it involves the activation of two serine-threonine Rho-regulated protein kinases, ROCK and PRK, which in turn regulate the contraction of acto-myosin fibers and the organization of focal adhesions, respectively. Ultimately, both kinase-initiated biochemical routes cooperate to promote cell-cell junction remodeling and loss of endothelial monolayer integrity.

Detailed genetic studies have revealed that thrombin activates RhoA through $G\alpha_{12/13}$ and to a lesser extend $G\alpha_{q/11}$ (9,42,43). Here, by using siRNA knockdown approaches, we show that in endothelial cells the expression of both $G\alpha_{12}$ and $G\alpha_{13}$ is required for the full activation of RhoA. In addition, $G\alpha_q$ and $G\alpha_{11}$ knockdown prevented RhoA activation by thrombin. This suggests that in endothelial cells the two families of heterotrimeric G protein α subunits, $G\alpha_{12/13}$ and $G\alpha_{11/q}$, play a complementary and non-redundant role. In the case of the $G\alpha_{12/13}$ family, they likely activate RhoA through guanine nucleotide exchange factor (GEF), such as p115-RhoGEF, PDZ-RhoGEF and leukemia-associated Rho GEF (LARG), which all exhibit a structural domain highly related to that of regulators of G-protein signaling (RGS), whereby they bind $G\alpha_{12/13}$ directly, leading to RhoA activation (44-47). In the case of $G\alpha_{11/q}$, this G protein family may utilize GEFs such as p63 RhoGEF and Trio (48,49), though it may also utilize LARG (50). Although which specific GEFs are in-volved in the RhoA activation by thrombin, and specifically how calcium elevation affects their activity or that of Rho GAPs (GTPase activating proteins), are yet to be determined. Nonetheless, the fact that coupling to $G\alpha_{12/13}$ and $G\alpha_{11/q}$ are both required, suggests a dual, fine-tuned regulation of RhoA activation downstream of PAR stimulation.

On the other hand, initial experiments using overexpression of RhoA mutants provided evidence of a role of its cycling activity in thrombin-induced endothelial permeability (10,14). Combined with the use of Y27632 as a ROCK pharmacologic inhibitor, it was suggested that the Rho/ROCK nexus affects both actin stress fiber and endothelial junction organization in response to thrombin (10,14). In this regard, Y27632 can specifically inhibit the kinase activity of two Rho-downstream effectors, ROCK and PRK (34). Indeed, by using siRNA approaches and mutant overexpression, we confirmed the involvement of ROCK, and revealed that PRK is an essential component of the signaling pathway deployed by thrombin to increase endothelial permeability. Future works are still required to assess whether PRK-KD mutant might have similar effects than the abolition of its expression by siRNA, as multiple domains within PRK protein, beside its kinase activity, might control downstream signaling events. In fact, we show here that the thrombin-induced signaling bifurcates downstream from RhoA, and that ROCK and PRK contribute by complementary mechanisms to trigger actin stress fiber formation associated with focal adhesions (Fig. 6). While MLC function and therefore the regulation of the actomyosin function are the most likely targets of ROCK in the mechanism by which thrombin induced stress fiber formation and cell contraction (13,35,51), PRK, also known as PKN, is a poorly studied but highly abundant RhoA target (52). PRK has been shown to play a role in actin and focal contact organization in response to lysophosphatidic acid (53,54). As speculated from the structural resemblance between the catalytic domain of PRK and PKC, PRK efficiently phosphorylates peptide substrates based on the pseudosubstrate sequence of PKC *in vitro*. Similarly, to the use of Y27632 that blocks both ROCK and PRK, the treatment with PKC inhibitors may have masked PRK implication in thrombin signaling as well (34,41). In the context of thrombin-induced endothelial permeability, our data strongly suggest that PRK is involved in the regulation of focal adhesion components. Indeed, the absence of PRK expression specifically hampered FAK and paxillin phosphorylation, and altered the morphology of focal adhesions. However, the direct target of PRK kinase activity is still to be determined. One can speculate that the actin-binding protein, α-actinin, might link PRK to focal adhesion in the thrombin-induced cytoskeletal changes in endothelial cells, as α-actinin has been shown to interact with PRK (55). In addition to the control of PRK activity by RhoA binding (56,57), phosphoinositide 3-kinase and PDK1 signaling might contribute to PRK activation (58,59), therefore providing a likely explanation for the inhibitory activity of phosphoinositide 3-kinase inhibitors on thrombin-induced signaling ((41) and data not shown).

Thrombin triggers the rapid contraction of the endothelial monolayer (35,41). This acto-myosin contractility increase is suspected to regulate in turn cellular adhesion at the level of cell-to-extracellular matrix and cell-to-cell contacts (15,41); this contraction appears to be the driven force for cell-cell contacts disengagement and loss of the endothelial barrier integrity. Alternatively, acto-myosin contraction could trigger not only mechanical and cytoskeletal changes in endothelial cells, but could also lead to the redistribution of kinases and phosphatases within the cells. This mechanism might participate in the phosphorylation of VE-cadherin and associated proteins in response to thrombin; such phosphorylation cascade is thought to disrupt cadherin-based cell-cell contacts (6,7,30,39,40). Of note, serine-dependent VE-cadherin internalization, which plays a central role in VEGF-induced vascular leakage (3,30), was slightly enhanced in thrombin-stimulated cells, suggesting the existence of divergent mechanisms between VEGF and thrombin in the regulation of the endothelial barrier.

In conclusion, while VEGF may promote endothelial cell permeability by a more linear signaling axis, the biochemical route by which thrombin induces the rapid loss of endothelial barrier integrity appears to involve a signaling network that involves multiple G proteins and their linked GEFs, which converges to activate RhoA, and that RhoA utilizes divergent molecular targets to promote the contraction of acto-myosin and the formation and remodeling of focal contacts. Ultimately, this signaling network may enable thrombin receptors to orchestrate rapid and reversible cytoskeletal changes and the reorganization of cell-matrix and cell-cell adhesion systems. Both mechanisms are required to facilitate junctional permeability and the rapid extravasation of serum components, thereby initiating the production of provisional matrix, vascular wall thickening, and a pro-inflammatory phenotype.

## REFERENCES

1. Red-Horse, K., Crawford, Y., Shojaei, F., and Ferrara, N. (2007) *Dev Cell* **12**(2), 181-194
2. Weis, S. M., and Cheresh, D. A. (2005) *Nature* **437**(7058), 497-504
3. Gavard, J., and Gutkind, J. S. (2006) *Nat Cell Biol* **8**(11), 1223-1234
4. Eliceiri, B. P., Paul, R., Schwartzberg, P. L., Hood, J. D., Leng, J., and Cheresh, D. A. (1999) *Mol Cell* **4**(6), 915-924
5. Stockton, R. A., Schaefer, E., and Schwartz, M. A. (2004) *J Biol Chem* **279**(45), 46621-46630
6. Weis, S., Shintani, S., Weber, A., Kirchmair, R., Wood, M., Cravens, A., McSharry, H., Iwakura, A., Yoon, Y. S., Himes, N., Burstein, D., Doukas, J., Soll, R., Losordo, D., and Cheresh, D. (2004) *J Clin Invest* **113**(6), 885-894
7. Dejana, E. (2004) *Nat Rev Mol Cell Biol* **5**(4), 261-270
8. Coughlin, S. R. (2005) *J Thromb Haemost* **3**(8), 1800-1814
9. Komarova, Y. A., Mehta, D., and Malik, A. B. (2007) *Sci STKE* **2007**(412), re8
10. Wojciak-Stothard, B., Potempa, S., Eichholtz, T., and Ridley, A. J. (2001) *J Cell Sci* **114**(Pt 7), 1343-1355
11. Vouret-Craviari, V., Grall, D., and Van Obberghen-Schilling, E. (2003) *J Thromb Haemost* **1**(5), 1103-1111
12. Klarenbach, S. W., Chipiuk, A., Nelson, R. C., Hollenberg, M. D., and Murray, A. G. (2003) *Circ Res* **92**(3), 272-278
13. Goeckeler, Z. M., and Wysolmerski, R. B. (2005) *J Biol Chem* **280**(38), 33083-33095
14. Essler, M., Amano, M., Kruse, H. J., Kaibuchi, K., Weber, P. C., and Aepfelbacher, M. (1998) *J Biol Chem* **273**(34), 21867-21874
15. Moy, A. B., Blackwell, K., and Kamath, A. (2002) *Am J Physiol Heart Circ Physiol* **282**(1), H21-29
16. Singh, I., Knezevic, N., Ahmmed, G. U., Kini, V., Malik, A. B., and Mehta, D. (2007) *J Biol Chem* **282**(11), 7833-7843
17. Mehta, D., Ahmmed, G. U., Paria, B. C., Holinstat, M., Voyno-Yasenetskaya, T., Tiruppathi, C., Minshall, R. D., and Malik, A. B. (2003) *J Biol Chem* **278**(35), 33492-33500
18. Mehta, D., Rahman, A., and Malik, A. B. (2001) *J Biol Chem* **276**(25), 22614-22620
19. van Nieuw Amerongen, G. P., van Delft, S., Vermeer, M. A., Collard, J. G., and van Hinsbergh, V. W. (2000) *Circ Res* **87**(4), 335-340
20. van Nieuw Amerongen, G. P., Draijer, R., Vermeer, M. A., and van Hinsbergh, V. W. (1998) *Circ Res* **83**(11), 1115-1123
21. Marinissen, M. J., Servitja, J. M., Offermanns, S., Simon, M. I., and Gutkind, J. S. (2003) *J Biol Chem* **278**(47), 46814-46825
22. Li, X., Hahn, C. N., Parsons, M., Drew, J., Vadas, M. A., and Gamble, J. R. (2004) *Blood* **104**(6), 1716-1724
23. Gorovoy, M., Niu, J., Bernard, O., Profirovic, J., Minshall, R., Neamu, R., and Voyno-Yasenetskaya, T. (2005) *J Biol Chem* **280**(28), 26533-26542
24. Birukova, A. A., Birukov, K. G., Smurova, K., Adyshev, D., Kaibuchi, K., Alieva, I., Garcia, J. G., and Verin, A. D. (2004) *Faseb J* **18**(15), 1879-1890
25. Takeya, R., Taniguchi, K., Narumiya, S., and Sumimoto, H. (2008) *Embo J* **27**(4), 618-628
26. Maekawa, M., Ishizaki, T., Boku, S., Watanabe, N., Fujita, A., Iwamatsu, A., Obinata, T., Ohashi, K., Mizuno, K., and Narumiya, S. (1999) *Science* **285**(5429), 895-898
27. Edgell, C. J., McDonald, C. C., and Graham, J. B. (1983) *Proc Natl Acad Sci U S A* **80**(12), 3734-3737
28. Marinissen, M. J., Chiariello, M., Tanos, T., Bernard, O., Narumiya, S., and Gutkind, J. S. (2004) *Mol Cell* **14**(1), 29-41
29. Kim, Y. M., and Benovic, J. L. (2002) *J Biol Chem* **277**(34), 30760-30768
30. Gavard, J., Patel, V., and Gutkind, J. S. (2008) *Dev Cell* **14**(1), 25-36

31. Xiao, K., Allison, D. F., Buckley, K. M., Kottke, M. D., Vincent, P. A., Faundez, V., and Kowalczyk, A. P. (2003) *J Cell Biol* **163**(3), 535-545
32. Ui, M., and Katada, T. (1990) *Adv Second Messenger Phosphoprotein Res.* **24**, 63-69
33. Ueda, S., Lee, S. L., and Fanburg, B. L. (1990) *Circ Res* **66**(4), 957-967
34. Davies, S. P., Reddy, H., Caivano, M., and Cohen, P. (2000) *Biochem J* **351**(Pt 1), 95-105
35. Kolodney, M. S., and Wysolmerski, R. B. (1992) *J Cell Biol* **117**(1), 73-82
36. Basile, J. R., Gavard, J., and Gutkind, J. S. (2007) *J Biol Chem* **282**(48), 34888-34895
37. Xiao, K., Garner, J., Buckley, K. M., Vincent, P. A., Chiasson, C. M., Dejana, E., Faundez, V., and Kowalczyk, A. P. (2005) *Mol Biol Cell* **16**(11), 5141-5151
38. Tan, W., Palmby, T. R., Gavard, J., Amornphimoltham, P., Zheng, Y., and Gutkind, J. S. (2008) *Faseb J* **22** (6), 1829-1838
39. Potter, M. D., Barbero, S., and Cheresh, D. A. (2005) *J Biol Chem* **280**(36), 31906-31912
40. Esser, S., Lampugnani, M. G., Corada, M., Dejana, E., and Risau, W. (1998) *J Cell Sci* **111 ( Pt 13)**, 1853-1865
41. Mehta, D., and Malik, A. B. (2006) *Physiol Rev* **86**(1), 279-367
42. Vogt, S., Grosse, R., Schultz, G., and Offermanns, S. (2003) *J Biol Chem* **278**(31), 28743-28749
43. Wirth, A., Benyo, Z., Lukasova, M., Leutgeb, B., Wettschureck, N., Gorbey, S., Orsy, P., Horvath, B., Maser-Gluth, C., Greiner, E., Lemmer, B., Schutz, G., Gutkind, S., and Offermanns, S. (2008) *Nat Med* **14**(1), 64-68
44. Fukuhara, S., Murga, C., Zohar, M., Igishi, T., and Gutkind, J. S. (1999) *J Biol Chem* **274**(9), 5868-5879
45. Fukuhara, S., Chikumi, H., and Gutkind, J. S. (2000) *FEBS Lett* **485**(2-3), 183-188
46. Hart, M. J., Jiang, X., Kozasa, T., Roscoe, W., Singer, W. D., Gilman, A. G., Sternweis, P. C., and Bollag, G. (1998) *Science* **280**(5372), 2112-2114
47. Kozasa, T., Jiang, X., Hart, M. J., Sternweis, P. M., Singer, W. D., Gilman, A. G., Bollag, G., and Sternweis, P. C. (1998) *Science* **280**(5372), 2109-2111
48. Rojas, R. J., Yohe, M. E., Gershburg, S., Kawano, T., Kozasa, T., and Sondek, J. (2007) *J Biol Chem* **282**(40), 29201-29210
49. Williams, S. L., Lutz, S., Charlie, N. K., Vettel, C., Ailion, M., Coco, C., Tesmer, J. J. G., Jorgensen, E. M., Wieland, T., and Miller, K. G. (2007) *Genes Dev.* **21**(21), 2731-2746
50. Booden, M. A., Siderovski, D. P., and Der, C. J. (2002) *Mol Cell Biol* **22**(12), 4053-4061
51. Kimura, K., Ito, M., Amano, M., Chihara, K., Fukata, Y., Nakafuku, M., Yamamori, B., Feng, J., Nakano, T., Okawa, K., Iwamatsu, A., and Kaibuchi, K. (1996) *Science* **273**(5272), 245-248
52. Palmer, R. H., Ridden, J., and Parker, P. J. (1994) *FEBS Lett* **356**(1), 5-8
53. Amano, M., Mukai, H., Ono, Y., Chihara, K., Matsui, T., Hamajima, Y., Okawa, K., Iwamatsu, A., and Kaibuchi, K. (1996) *Science* **271**(5249), 648-650
54. Watanabe, G., Saito, Y., Madaule, P., Ishizaki, T., Fujisawa, K., Morii, N., Mukai, H., Ono, Y., Kakizuka, A., and Narumiya, S. (1996) *Science* **271**(5249), 645-648
55. Mukai, H., Toshimori, M., Shibata, H., Takanaga, H., Kitagawa, M., Miyahara, M., Shimakawa, M., and Ono, Y. (1997) *J Biol Chem* **272**(8), 4740-4746
56. Maesaki, R., Ihara, K., Shimizu, T., Kuroda, S., Kaibuchi, K., and Hakoshima, T. (1999) *Mol Cell* **4**(5), 793-803
57. Mukai, H. (2003) *J Biochem* **133**(1), 17-27
58. Flynn, P., Mellor, H., Casamassima, A., and Parker, P. J. (2000) *J Biol Chem* **275**(15), 11064-11070
59. Dong, L. Q., Landa, L. R., Wick, M. J., Zhu, L., Mukai, H., Ono, Y., and Liu, F. (2000) *Proc Natl Acad Sci* **97**(10), 5089-5094

**FOOTNOTE**

This research was supported by the Intramural Research Program of NIH, National Institute of Dental and Craniofacial Research, the Centre National de la Recherche Scientifique (CNRS), and the Institut National de la Santé et de la Recherche Médicale (INSERM). The authors declare that they have no competing financial interests.

The abbreviations used are: VEGF, vascular endothelial growth factor; PAK, p21-activated kinase; PAR, protease activated receptor; MLC, myosin light chain kinase; PRK, PKC-related kinase; FAK, focal adhesion kinase; ERK, extracellular related kinase; KD, kinase dead; G-protein Exchange factor, GEF.

## FIGURE LEGENDS

Fig. 1. Thrombin induces endothelial permeability through PAR-1 and $G\alpha_{12/13}$ and $G\alpha_{11/q}$ heterotrimeric G protein subunits.
*A*. Human endothelial cells were cultivated on collagen-coated 3 µm pore-size inserts for 3 days, starved overnight and pre-treated with PAR antagonists ($PAR^{inh}$), for 4 hours at 1 µM (PAR1), 20 µM (PAR2) and 10 µM (PAR 3/4). Cells were then stimulated with thrombin (0.1 U/ml, 15 min) and incubated with 40 kDa FITC-dextran for 30 minutes more. Results are shown as the average $\pm$ S.E.M. of the ratio between treated and untreated cells in 3 independent experiments. *B*. Total cell lysates were extracted from similarly treated cells and subjected to phospho-ERK1/2 and total ERK1 western-blots. *C*. As described in *A*, endothelial monolayers were incubated with pertussis toxin (PTX, 10 ng/ml) overnight before stimulation and permeability assay. Cells were stimulated with VEGF (50 ng/ml, 30 min), thrombin (0.1 U/ml, 15 min), and IL-8 (50 ng/ml, 30 min). *D*. Endothelial cells were electroporated with $G\alpha_{12}$, $G\alpha_{13}$, $G\alpha_{q}$ and $G\alpha_{11}$, either wilt-type (WT) or constitutively active mutant (QL) and cell lysates were prepared 48 hours later. *E-F*. Similarly transfected cells were processed for permeability assays in non stimulated cells in *E*, or stimulated by VEGF (50 ng/ml, 30 min) or thrombin (0.1 U/ml, 15 min) in *F*. Bar graphs show the average $\pm$ S.E.M. of the ratio between QL transfected cells and WT transfected cells for each $G\alpha$ subunits. *G*. Endothelial cells were transfected with non-silencing (nsi) or siRNA sequences (50 nM, 5 days) targeting $G\alpha_{12}$, $G\alpha_{13}$, $G\alpha_{q}$ (2 independent sequences) and $G\alpha_{11}$ (2 independent sequences). Protein lysates were analyzed for $G\alpha_{12}$, $G\alpha_{13}$, $G\alpha_{q}$ and $G\alpha_{11}$ expression while Tubulin was used as a loading control. *H*. Similarly siRNA-treated cells were used for permeability assays in response to thrombin. All panels are representative of at least 3 independent experiments. ANOVA test: *, $p<0.05$; **, $p<0.01$; ***, $p<0.001$.

Fig. 2. $G\alpha_{11/q}$-induced calcium entry is required for Thrombin signaling in permeability.
*A*. Intracellular calcium concentration was measured by the Fluo4 probe in overnight starved 3 days-old endothelial monolayers and left unstimulated (buffer) or stimulated by ionomycin (1µM), thrombin (0.1 U/ml) and VEGF (50 ng/ml) for the indicated times. *B*. Similar experiments were performed in non-silencing (nsi) and $G\alpha_{12/13}$, $G\alpha_{q}$, $G\alpha_{11}$ and $G\alpha_{q/11}$ siRNA-treated cells. *C*. Human endothelial cells were cultivated on collagen-coated 3 µm pore-size inserts for 3 days, starved overnight and pre-treated with the calcium chelator (BAPTA-AM, 25 µM), the inhibitors of storage-operated calcium ($SOC^{inh}$) flux (2-ABP, 75 µM and TMB-8, 100 µM), and the inhibitors of receptor-operated calcium ($ROC^{inh}$) flux (lanthanade, 250 nM and verapamil, 500 nM) for 30 min prior to thrombin stimulation and permeability assays. *D*. Human endothelial cells were transfected with non-silencing RNA (nsi) or $G\alpha_{12/13}$, $G\alpha_{q/11}$ and $G\alpha_{12/13/q/11}$ (all 4) siRNA for 5 days. Overnight starved cells were non-stimulated (-) or stimulated by thrombin, and then subjected to RhoA pull-down assay to assess RhoA activation (RhoA-GTP). Western-blots were performed in total cell lysates for the indicated G protein $\alpha$ subunits and RhoA. The level of RhoA-GTP was estimated on scanned membranes (Licor) and expressed as a normalized ratio on total RhoA. *E*. Overnight starved 3 days-old endothelial cells were treated with 25 µM BAPTA-AM (bapta), 75 µM 2-ABP, and 250 nM lanthanade (lantha) prior to thrombin stimulation. RhoA activation was estimated in the Rho pull-down fraction (RhoA-GTP), and PRK and MLC phosphorylations (p) in total cell lysates. Total RhoA served as a loading control. All panels are representative of at least 3 independent experiments. ANOVA test: *, $p<0.05$; **, $p<0.01$; ***, $p<0.001$.

Fig. 3. ROCK and PRK are both required for Thrombin-induced permeability.
*A-B*. Permeability assays were done in 3 days-old human endothelial cell starved overnight and pre-treated with C3 toxin (1 µg/ml) with tetanolysin (20 units) for 16h, Y27632 (45 min, 5, 10 and 50 nM), blebbistatin (45 min, 5 µM) and ML-7 (45 min, 10 µM) prior VEGF and thrombin stimulation as described previously. *C*. Human endothelial cells were electroporated with mock or myc-tagged ROCK-KD mutant and cultured for 3 days on collagen-coated inserts. Overnight starved cells were then stimulated with thrombin and further analyzed for permeability. Expression level of ROCK-KD was checked by western-blot against myc. *D.* Similarly, cells were transfected with non-silencing RNA (nsi, 3 days, 50 nM) or two independent sequences targeting PRK (si#1 and si#2, 3 days, 50 nM). Extinction efficiency was checked by western-blot for PRK, and Tubulin was used as a loading control. All panels are representative of at least 3 independent experiments. Permeability assays were performed in each condition in non-stimulated (Ctrl), and VEGF and Thrombin-stimulated cells. ANOVA test: *, $p<0.05$; **, $p<0.01$; ***, $p<0.001$.

Fig. 4. ROCK and PRK control distinct Thrombin-induced contractility pathways.
*A.* Human endothelial cells were transfected with 50 nM non-silencing RNA (nsi) or two independent sequences targeting PRK (si#1 and si#2) for 3 days. Overnight starved cells were left unstimulated or simulated with thrombin (0.1 U/ml, 5 min) and total cell lysates were used for western-blot against phospho-specific (p) MLC, cofilin (Cof), paxillin (Pax), and FAK. siRNA knockdown was checked by PRK western-blot, Tubulin was used as a loading control. *B.* Similar experiments were performed in endothelial cells transfected with mock DNA or myc-tagged ROCK-KD, whose expression level was checked by myc western-blots. *C.* The levels of phosphorylation of MLC, cofilin, (Cof), paxillin (Pax) and FAK were estimated on scanned membranes (Licor) from multiple experiments and expressed as a normalized ratio on tubulin. *D.* Nsi-, PRK siRNA-, mock- and ROCK-KD-transfected human endothelial cells were grown on collagen-coated glass coverslips for 3 days as monolayers, starved overnight and then stimulated with thrombin. Non-transfected cells were left unstimulated and used as control (Ctrl). Actin cystoskeleton (red) was revealed by phalloidin labeling, focal adhesions (green) by anti-paxillin immunostaining. Binary calculation (bin.) was applied on paxillin threshold pictures to unveil focal adhesion morphology (Image J software). *E.* The number of cells presenting actin stress fibers was manually counted in 250 cells and expressed as the percentage of total cells in non-stimulated (ctrl) and nsi-, PRK siRNA-, mock- and ROCK-KD-transfected cells exposed to Thrombin. *F-G.* Similarly treated cells were analyzed by confocal microscopy and post-treated for binary processing on threshold pictures. Both the number of particles (*F*) and the average pixel size (*G*) were calculated for the paxillin-positive focal adhesions per field (n=9). All panels are representative of at least 3 independent experiments. ANOVA test: *, $p<0.05$; **, $p<0.01$.

Fig. 5. ROCK- and PRK-induced contractility is involved in cell-cell contact remodeling.
*A.* Human endothelial cells were grown on collagen-coated glass coverslips for 3 days as monolayers, starved overnight and then stimulated with thrombin and VEGF. Cell-cell junctions (green) were revealed by either VE-cadherin or ZO-1 immunostaining. Nuclei were counterstained by DAPI (magenta). Binary calculation was applied on threshold pictures to unveil cell-cell morphology (Image J software). Individual cells are numerated 1 to 4. *B.* Human endothelial cells were grown on collagen-coated glass coverslips for 3 days as monolayers, starved overnight, incubated with anti-VE-cadherin at 4°C for 1h, and then stimulated with thrombin and VEGF at 37°C for 30 min. Cells were subjected to a mild acid-wash prior fixation and immunofluorescence. Cells with internal acid-resistant staining for VE-cadherin were scored positive when they displayed at least 5 vesicle-like internal staining. Graph represents the percentage of cells with VE-cadherin uptake in the total cell population. *C.* Human endothelial cells were grown for 3 days as monolayers, starved overnight and then stimulated with thrombin and VEGF. Total cell lysates (TCL) were blotted against phospho-S665 VE-cadherin and total VE-cadherin (VE-cad). Phospho-tyrosine (pY), β-catenin (β-cat), β-arrestin2 (β-arr2) and VE-cad western-blots were performed in VE-cad immunoprecipitation (IP) fractions. *D* mock-, PRK siRNA- and ROCK-KD-transfected human endothelial cells were grown on collagen-coated glass coverslips for 3 days as monolayers, starved overnight and then stimulated with thrombin. VE-cadherin immunostaining is shown in green. Nuclei were counterstained by DAPI (blue). Binary calculation (bin.) was applied on threshold pictures to unveil cell-cell morphology (Image J software). Individual cells are numerated 1 to 4. All panels are representative of at least 3 independent experiments. ANOVA test: *, $p<0.05$; **, $p<0.01$; ***, $p<0.001$.

Fig. 6. Signaling model and molecular mechanisms involved in Thrombin-induced permeability.
Thrombin induces the activation of its cognate receptor, PAR-1, which is coupled to $G\alpha_{11/q}$ and $G\alpha_{12/13}$ signaling. Knockdown experiments showed that $G\alpha_{11/q}$ is required for both calcium entry and RhoA activation upon throm-bin exposure, while Gα$_{12/13}$ cooperates in RhoA activation. This dual coupling might regulate the activity and localization and cycling of multiple Rho GEFs. RhoA is then central in the thrombin-induced increase of permeability in endothelial cells, as it triggers the activation of two essential downstream effectors, ROCK and PRK. In addition, RhoA can control formin activity on actin and microtubule organization. ROCK activity is required for acto-myosin contractility through the regulation of MLC and cofilin and the actin stress fiber architecture. Conversely, PRK expression contributes to FAK and paxillin phosphorylations and the proper focal adhesion morphology. Finally, both signaling routes converge on the increase of acto-myosin contractility and cell-cell junction remodeling, leading to endothelial barrier destabilization and permeability increase, by a distinct mechanism from that initiated by VEGF.

1. Thrombin increases endothelial permeability through Par-1 and Gα$_{12/13}$ and Gα$_{11/q}$ heterotrimeric G protein subunits

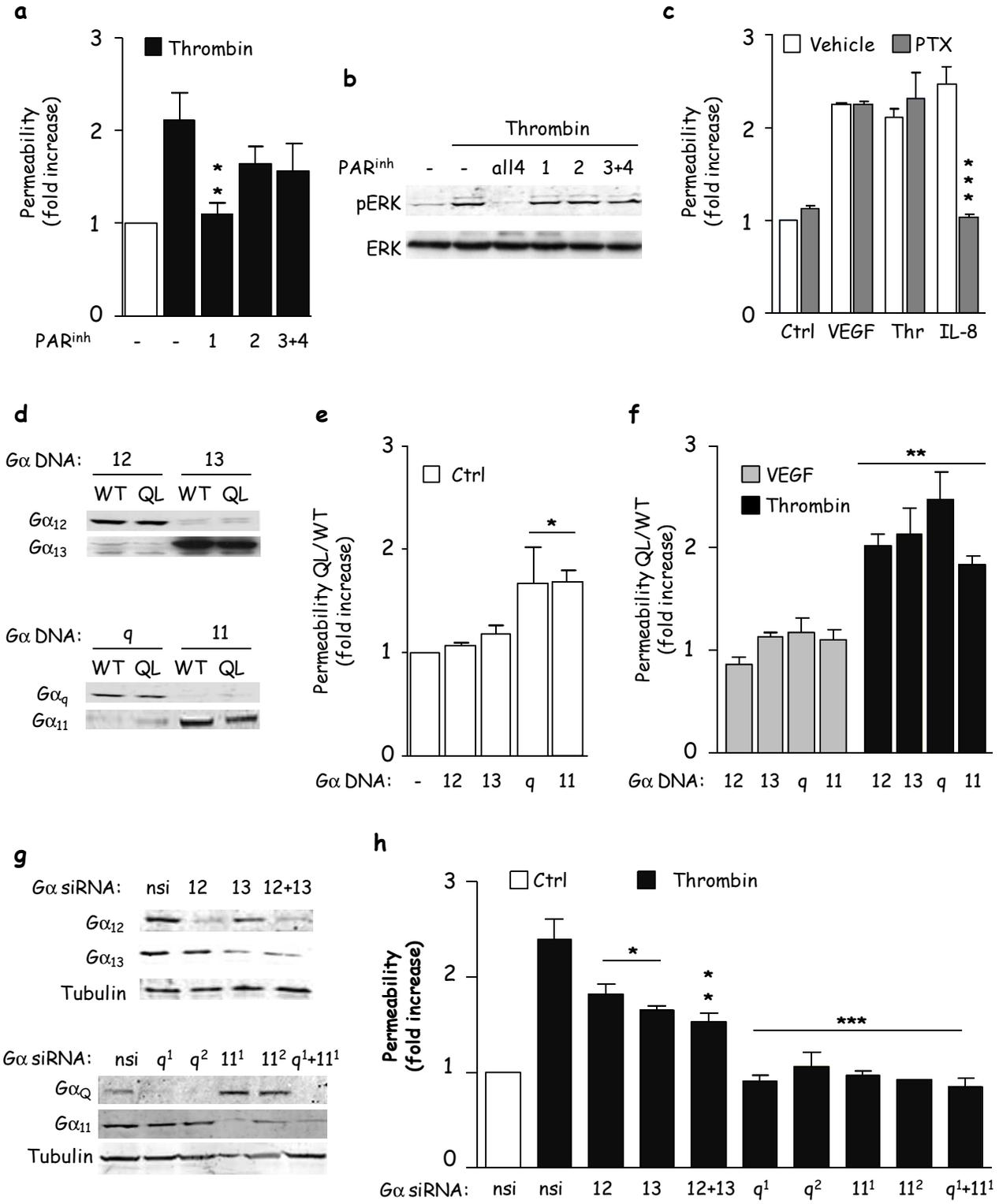

## 2. $G\alpha_{11/q}$-induced calcium entry is required for Thrombin signaling in permeability

a

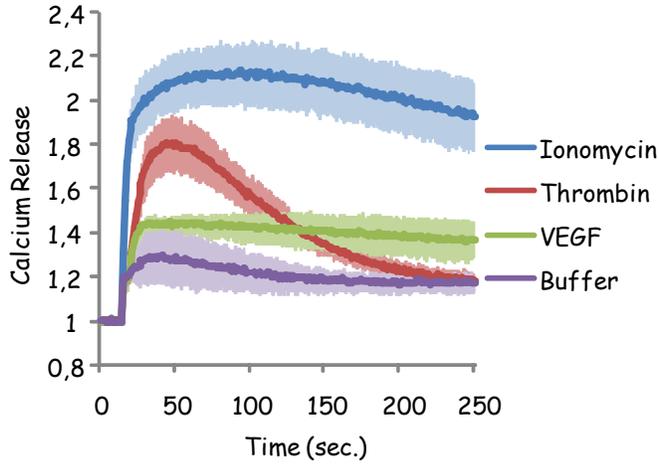

b

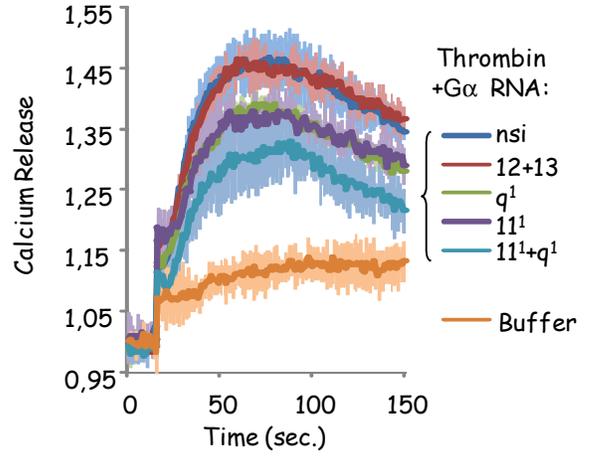

c

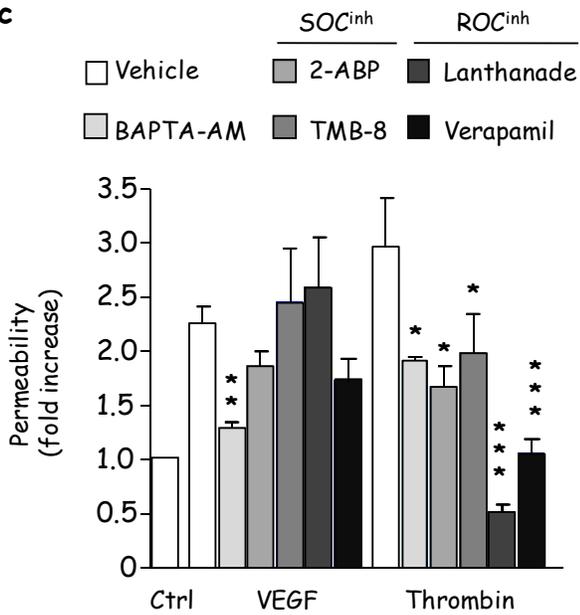

d

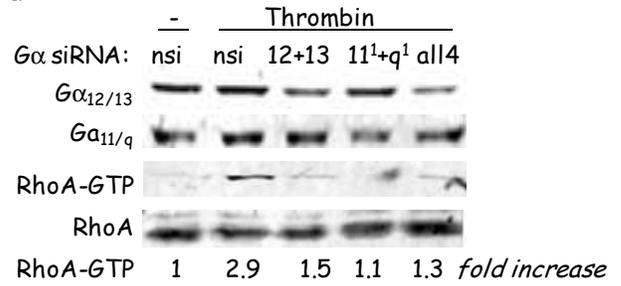

e

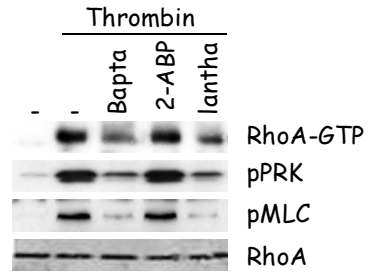

## 3. ROCK and PRK are both required for Thrombin-induced permeability

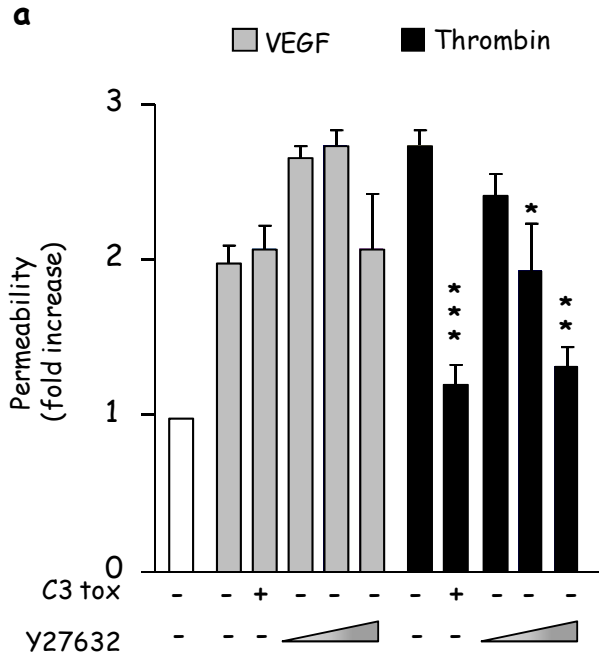
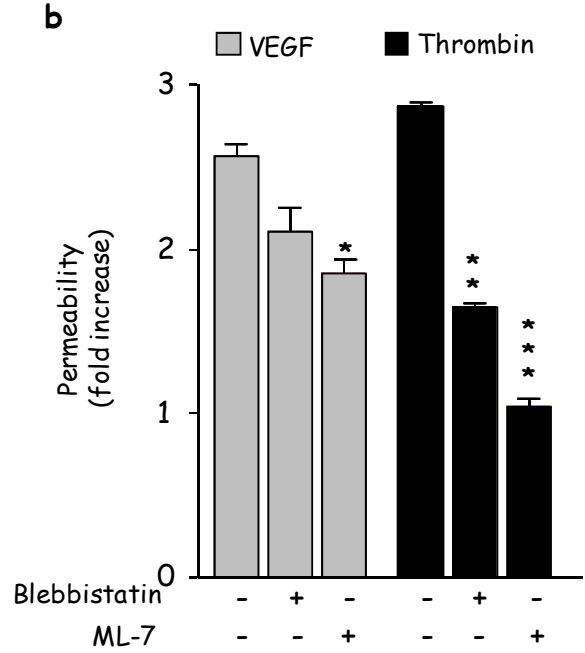
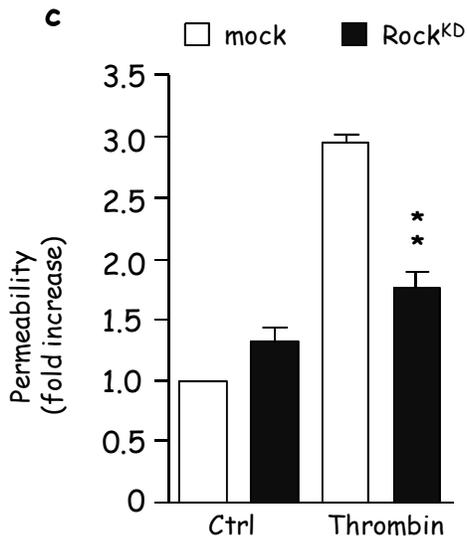
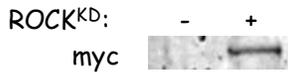
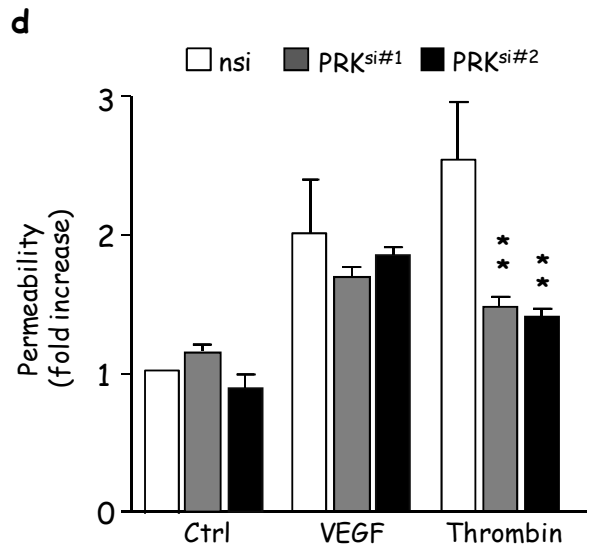
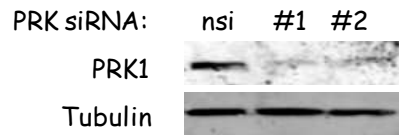

## 4. ROCK and PRK control separately Thrombin-induced contractility pathways

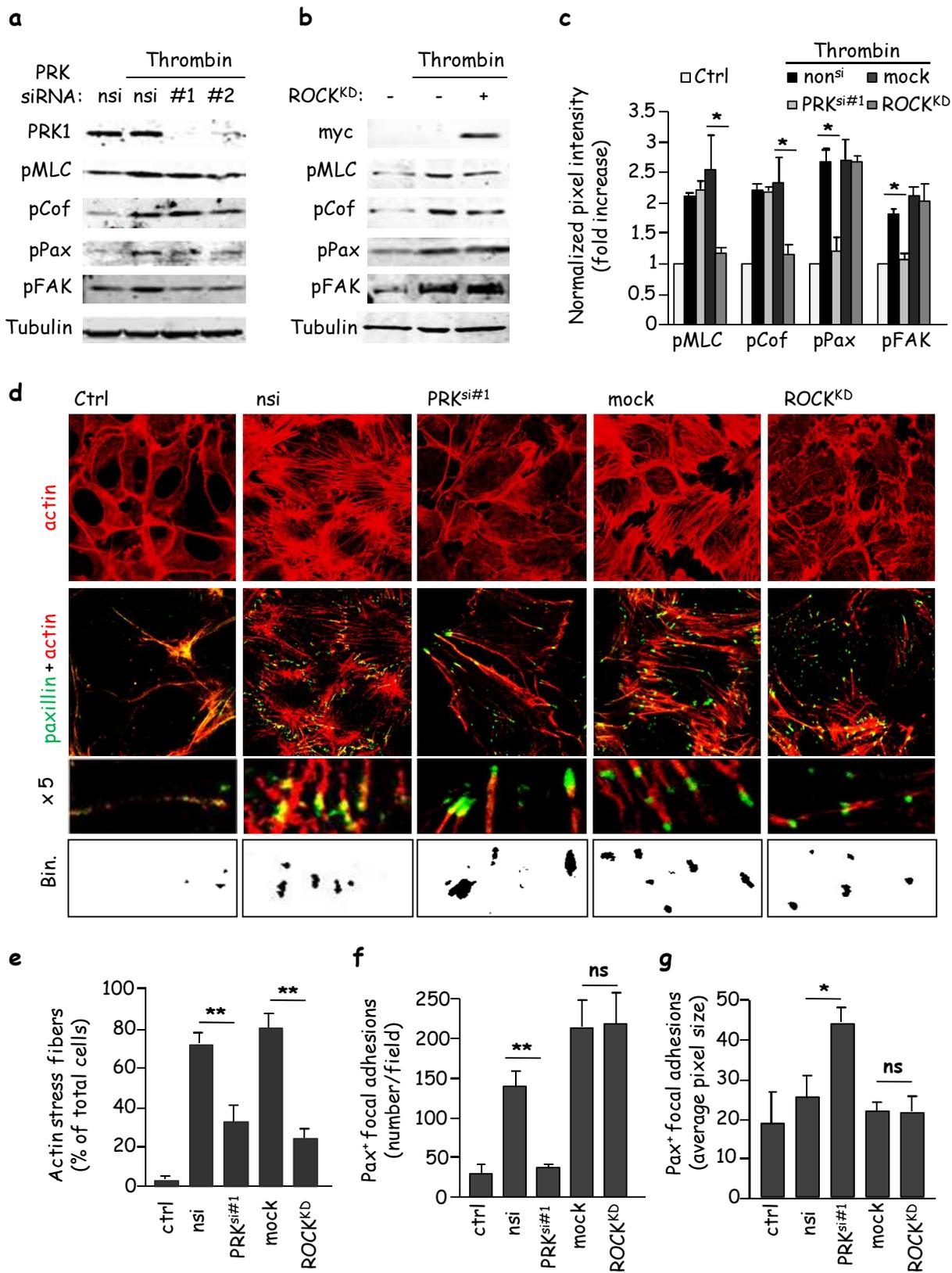

## 5. ROCK- and PRK-induced contractility is involved in cell-cell contact remodeling

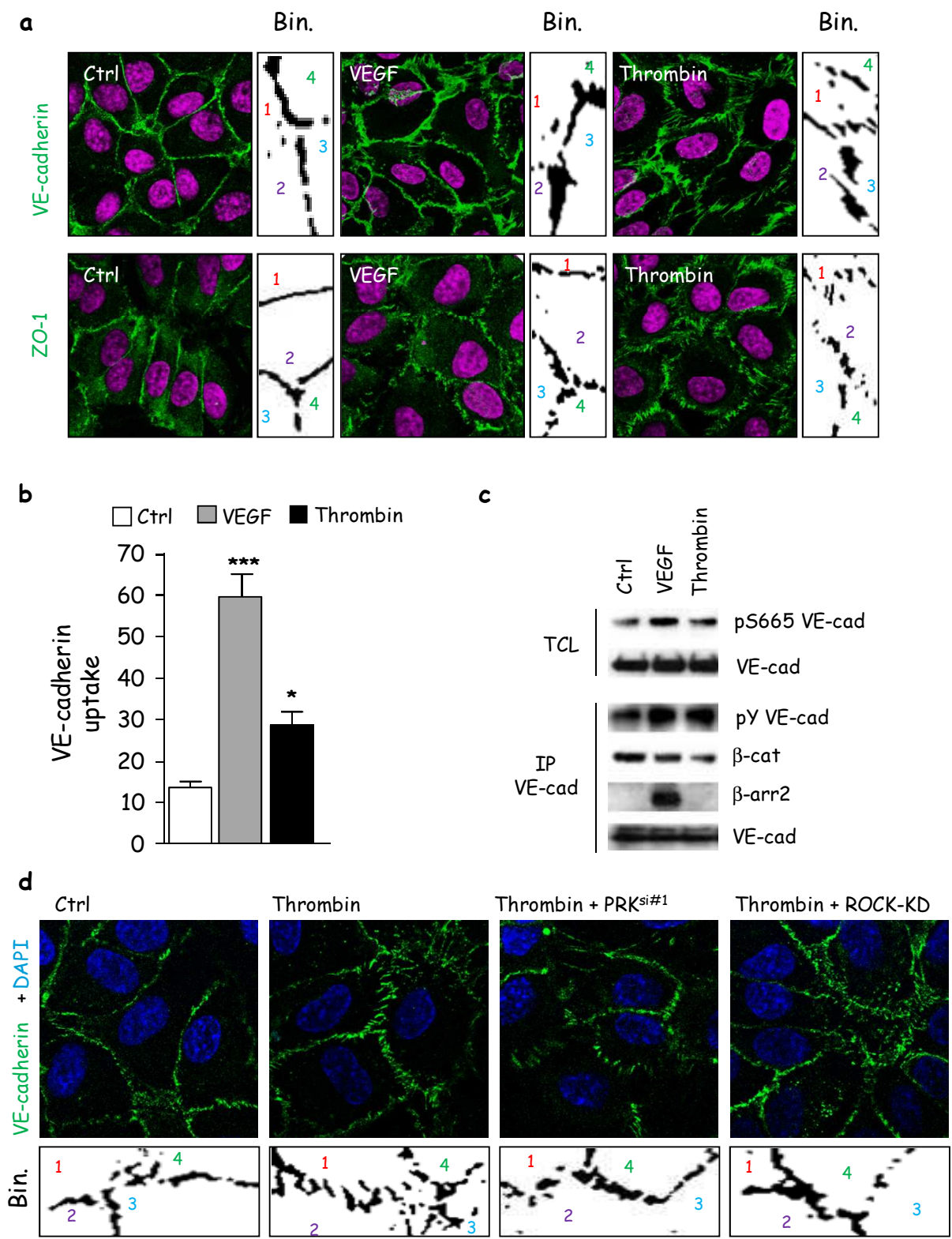

6. Signaling model and molecular mechanisms involved in Thrombin-induced permeability

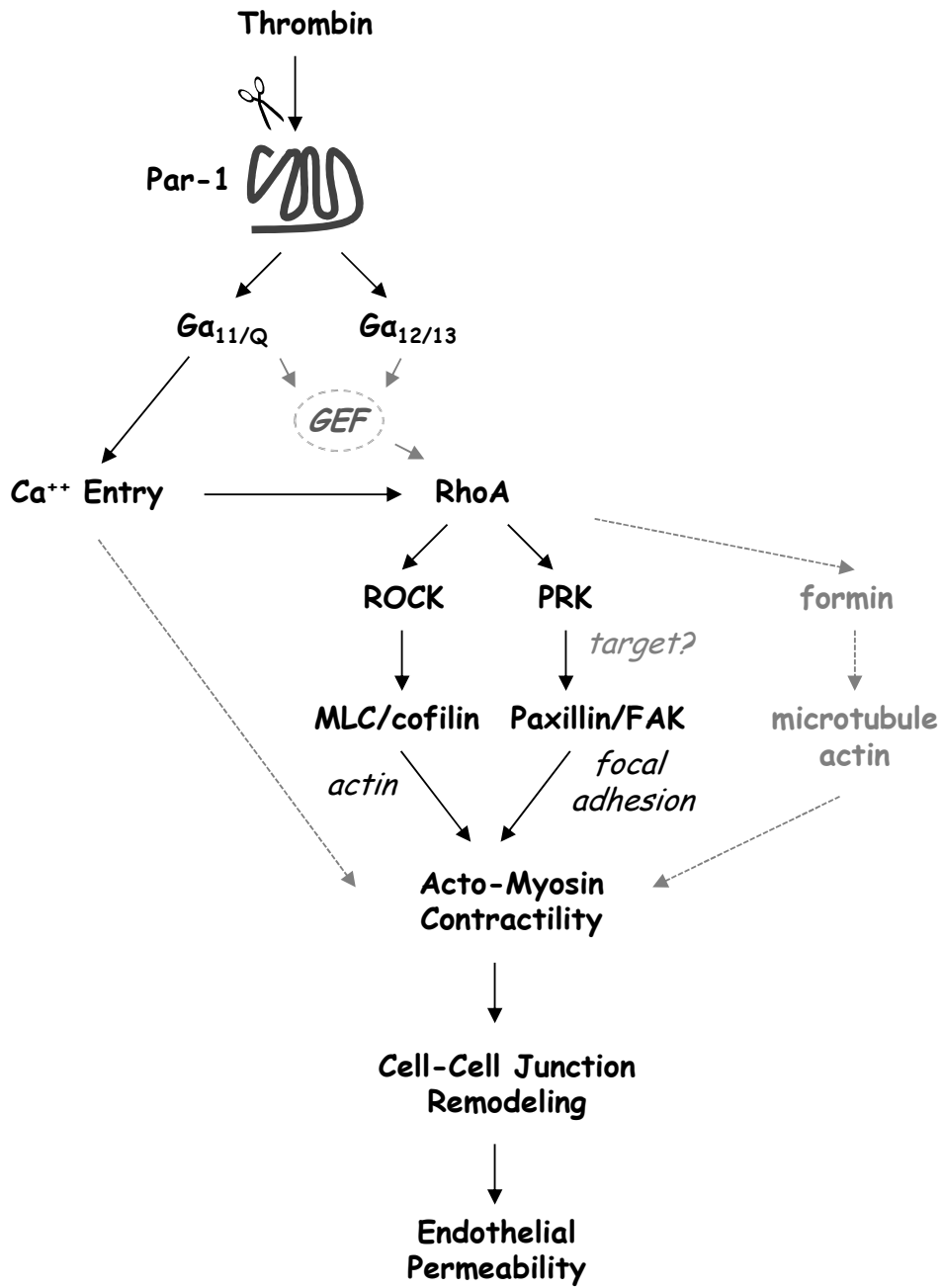